\documentclass[a4paper,fleqn]{cas-dc}

\usepackage[numbers]{natbib}
\usepackage{microtype}
\usepackage{graphicx}
\usepackage{dcolumn}
\usepackage{bm}
\graphicspath{ {./images/} }
\usepackage{color}
\usepackage{hyperref}
\usepackage{footnote}
\usepackage{wrapfig}
\usepackage{subcaption}
\usepackage[T1]{fontenc}
\usepackage[version=3]{mhchem}
\usepackage[T1]{fontenc}

\usepackage{amsmath}
\usepackage{amssymb}
\usepackage{amsthm}
\usepackage{amsfonts}
\usepackage{braket}
\usepackage{tabularx} 
\usepackage{tablefootnote}

\usepackage{mhchem}

\begin{document}

\shorttitle{Machine Learning for Improved DFT Thermodynamics}  
\shortauthors{Simak {\it et al.}}

\title [mode = title]{ Machine Learning for Improved Density Functional Theory Thermodynamics}

\author[1,2]{Sergei I. Simak} [orcid=0000-0002-1320-389X]
\credit{developed and conducted the analysis using supervised machine learning techniques} 
\author[2,3]{Erna K. Delczeg-Czirjak} [orcid=0000-0002-1667-2894]
\credit{carried out the density functional theory (DFT) calculations} 
\author[2,3]{Olle Eriksson} [orcid=0000-0001-5111-1374]
\credit{suggested the project and formulated the initial idea.
The authors collaboratively conceived the overall research concept and jointly authored the paper}
\address[1]{Department of Physics, Chemistry and Biology (IFM), Linköping University, SE-581 83 Linköping, Sweden}
\address[2]{Department of Physics and Astronomy, Uppsala University, Box 516, SE-75120, Uppsala, Sweden}
\address[3]{WISE - Wallenberg Initiative Materials Science for Sustainability, Department of Physics and Astronomy, Uppsala University, SE-751 20 Uppsala, Sweden}


\begin{abstract}
The predictive accuracy of density functional theory (DFT) for alloy formation enthalpies is often limited by intrinsic energy resolution errors, particularly in ternary phase stability calculations. In this work, we present a machine learning (ML) approach to systematically correct these errors, improving the reliability of first-principles predictions. A neural network model has been trained to predict the discrepancy between DFT-calculated and experimentally measured enthalpies for binary and ternary alloys and compounds. The model utilizes a structured feature set comprising elemental concentrations, atomic numbers, and interaction terms to capture key chemical and structural effects. By applying supervised learning and rigorous data curation we ensure a robust and physically meaningful correction. The model is implemented as a multi-layer perceptron (MLP) regressor with three hidden layers, optimized through leave-one-out cross-validation (LOOCV) and k-fold cross-validation to prevent overfitting. We illustrate the effectiveness of this method by applying it to the Al-Ni-Pd and Al-Ni-Ti systems, which are of interest for high-temperature applications in aerospace and protective coatings.

\end{abstract}



\begin{keywords}
Density Functional Theory (DFT),
Machine Learning Corrections,
Neural Network Regression,
Formation Enthalpy Prediction,
Phase Stability,
Multicomponent Alloys,
First-Principles Calculations,
High-Temperature Materials \sep 
\end{keywords}

\maketitle

\section{Introduction}
The ability to make reliable predictions of material properties using fast and accurate theoretical methods is highly desirable and is one of the main reasons for the widespread use of density functional theory (DFT) \cite{Kohn1965}. In several studies, experimental investigations have followed theoretical predictions of functional material properties, often resulting in joint publications. There are numerous notable successes where theoretical predictions preceded experimental verification.

A well-known example is the linear band dispersion of graphene, which was calculated using electronic structure theory \cite{Wallace1947} before being confirmed by angle-resolved photoemission spectroscopy (ARPES) experiments \cite{KATSNELSON200720}. Another example is the tunneling magnetoresistance (TMR) device, widely used in magnetic field sensor applications, which was predicted using first-principles electronic structure calculations \cite{Butler2001} before its experimental realization \cite{Bowen2001}.

More recently, interest has shifted toward magnetic materials in reduced dimensions, particularly thin two-dimensional (2D) systems. Notably, the class of materials Cr$_2$X$_2$Te$_6$ (where X = silicon or germanium), ZPS$_3$ (where Z = manganese, iron, or nickel), and iron telluride (FeTe) were all predicted by DFT-based calculations \cite{Lebegue2013} before their experimental confirmation \cite{Gong2019, D0NR06813F}.

Although several reviews \cite{borisov2023electronicstructuremagnetismskyrmions} have outlined in detail how density functional theory (DFT)-based calculations can accurately reproduce equilibrium volumes, elastic constants, structural stability, phonon frequencies, and magnetic properties of many materials, there are still important areas of materials science where DFT has not yet reached its full predictive potential. This limitation arises from the inherent accuracy of the energy functionals used in these calculations, which lack the necessary energy resolution.

\begin{figure}[!ht]
  \centering   \includegraphics[width=0.5\textwidth]{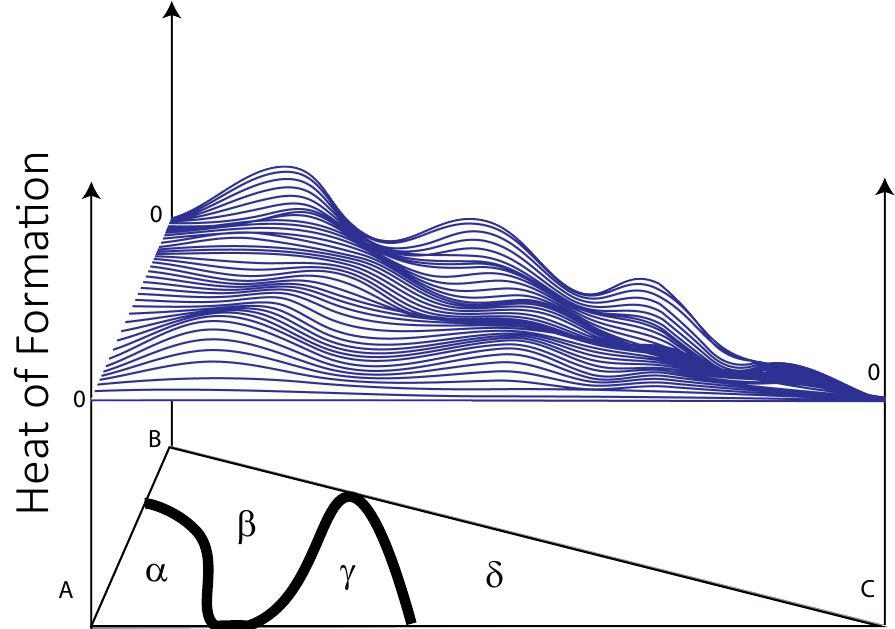}
    \caption{A schematic illustration of the energy landscape of the heat of formation for a ternary system composed of elements (or compounds) A, B, and C. Due to the peaks and valleys in the heat of formation landscape, the phases $\alpha$, $\beta$, $\gamma$, and $\delta$ are formed.}
    \label{ternary_phasediagram}
\end{figure}

One of the most significant challenges in this context is the ability of theoretical methods to predict the phase stability of compounds and alloys, particularly in the case of ternary phase diagrams, i.e., systems involving three elements. A schematic phase diagram is shown in Fig.~\ref{ternary_phasediagram}, where the symbols A, B, and C represent elements (or compounds) that form competing phases depending on concentration. The landscape of the heat of formation determines which phase should be stable (denoted as $\alpha$, $\beta$, $\gamma$, and $\delta$ in Fig.~\ref{ternary_phasediagram}).
Experimentally, such phase diagrams are published for most materials, although for some systems they may be incomplete. These diagrams serve as essential tools for identifying functional materials and understanding their physical and chemical properties. However, the direct application of DFT to predict complete phase diagrams, or even to reproduce known ones, is hindered by the intrinsic errors in DFT-calculated energies, which prevent accurate phase stability determinations.

Here, we illustrate this issue through a detailed investigation of two ternary phase diagrams: Al-Ni-Pd and Al-Ni-Ti. These systems are frequently studied for their potential to enhance the protective properties of nickel aluminides in high-temperature applications. Nickel aluminides serve as protective coatings against high-temperature oxidation and hot corrosion \cite{Aluminides1, aluminides2} in nickel-based superalloys used for aircraft engine turbine blades. Additionally, palladium-modified aluminide coatings function as bond coatings for thermal barrier systems, offering improved resistance to oxidation and hot corrosion \cite{Pd-modif1, Pd-modif2}. Ti-Al and Ti-Ni-Al alloys are also actively researched for aerospace applications due to their high strength, low density, and corrosion resistance at elevated operating temperatures \cite{SINA2016294, DRAPER2007330, SCHUSTER20061304}.

We outline an approach of reducing the error in density functional theory (DFT) for phase stability calculations. A simple linear correction, based on known enthalpy differences between DFT-calculated and experimentally measured values, provides a visible yet limited improvement. By applying machine learning techniques—specifically neural networks with supervised training—the predictive accuracy is significantly enhanced, enabling a more reliable determination of phase stability.

Our goal in this study is not to perform high-throughput calculations on thousands of materials with known experimental formation enthalpies to develop a universally applicable error correction model. Instead, we focus on demonstrating how such a model can be constructed and validating its predictive capability, even with a limited training dataset. This approach highlights the potential of machine learning for improving phase stability predictions while maintaining computational efficiency.

\section{Theoretical tools}
\subsection{Enthalpy of formation}
Though it is straightforward to calculate phase equilibria at given external conditions (temperature and pressure) through Gibbs or Helmholtz free energy calculations, such calculations can be complex and time-consuming at high temperatures due to contributions from phonons, anharmonic atomic vibrations, and other effects. Additionally, treating phonons in alloys requires a sophisticated approach. Therefore, in this paper, we focus on the ambient-temperature regions of phase diagrams, which requires only fast calculations without treatment of phonons (or magnons) that typically are done at 0 K.

To determine phase stability at ambient conditions, one needs the total energy of a specific phase as well as all competing phases that may form. We consider the simplest case prone to DFT errors relative to experimental values—namely, the enthalpy of formation ($\it H_f$) of each material. This enthalpy is determined from the DFT total energy relative to the most stable elemental structures as follows:
\begin{equation} \label{eq:H_f}
H_f (A_{x_A}B_{x_B}C_{x_C}\dots) = \begin{array}{l}
H(A_{x_A}B_{x_B}C_{x_C}\dots) - x_A H(A) \\- x_B H(B)
- x_C H(C) - \dots
\end{array}
\end{equation}
where $\it H(A_{x_{A}}B_{x_{B}}C_{x_{C}})$ 
is the enthalpy per atom of the intermetallic compound or alloy, and $\it H(A)$, $\it H(B)$, and $\it H(C)$ are the enthalpies per atom of the elements A, B and C in their ground-state structures. In this work we consider systems with maximum three elements, with A, B and C to be among the Al, Ni, Ti, and Pd. The ground-state structures of these elements are fcc-Al, fcc-Ni, fcc-Pd, and hcp-Ti, where fcc stands for face-centered cubic and hcp stands for hexagonal close-packed. Furthermore, $\it x_{A}$, $\it x_{B}$, and $\it x_{C}$=$1-x_A-{x}_{B}$ are the concentration of elements $\it A$, $\it B$, and $\it C$, respectively. When compared to experimental values of the enthalpy of formation, the error inherent in DFT based calculations of $H_f$ is unfortunately too large to enable a predictive capability to determine the relative stability of competing phases. It is the purpose of this work to outline a way to reduce this error. 

\subsection{Total energy calculations}
Total energies based on DFT are calculated using the exact muffin-tin orbital (EMTO) method \cite{Andersen1994,Vitos2007} in combination with the full charge density technique \cite{Kollar2000} at zero temperature and pressure and without zero-point motion. The chemical disorder is treated within the coherent potential approximation (CPA) \cite{Soven1967, Gyorffy1972} (EMTO-CPA \cite{Vitos2001}). The electrostatic correction to the single-site CPA is considered as implemented in the Lyngby version of the EMTO code \cite{Ruban2016}. For details, the reader is referred to Refs. \cite{Ruban2016}, \cite{screen1}, and \cite{screen2}. 
The one-electron Kohn-Sham equations are solved within the soft-core and scalar-relativistic approximations, with $l_{\rm max} = 3$ for partial waves and $l_{\rm max} = 5$ for their "tails". The Green's function is calculated for 16 complex energy points distributed exponentially on a semi-circular contour including states within 1 Ry below the Fermi level. The exchange-correlation effects are described within the Perdew-Burke-Ernzerhof \cite{PBE} version of the generalized gradient approximation. The 0 K theoretical equilibrium lattice parameter for each system is determined from a Morse type of equation of state \cite{Morse_eos} fitted to the {\it ab initio} total energies of the experimentally reported structures for five different atomic volumes. The heat of formation is calculated at the theoretical equilibrium volume of all systems used in Eqn.~\ref{eq:H_f}.
To ensure the convergence of total energy and volume calculations, the Monkhorst-Pack k-point mesh \cite{Monkhorst-Pack} is set to 17 $\times$ 17 $\times$ 17 within the irreducible wedge of the Brillouin zone for the cubic systems. For non-cubic structures, the k-point mesh is scaled according to the $b/a$ and $c/a$ ratios.

\subsection{Machine learning}
To improve the accuracy of first-principles calculations for a multicomponent compound or alloy formation enthalpies, we have developed a simple linear and a more involved neural network model, that are used here to predict the errors between computed and experimental enthalpies of formation for binary and ternary alloys. Each material is characterized using a structured set of input features, including elemental concentrations, atomic numbers, and interaction terms. A training dataset of reliable experimental values of the enthalphy of formation is initially filtered to exclude missing or unreliable enthalpy values, ensuring that only well-defined data points are used for training of the neural network. The input features are also normalized to prevent variations in scale from affecting model performance. The details of this are outlined below.

For a given material composed of elements \( A, B, C, \dots \), the elemental concentration vector is defined as:
\begin{equation} \label{eq:concentration}
\mathbf{x} = [x_A, x_B, x_C, \dots]
\end{equation}
where \( x_i \) represents the atomic fraction of element \( i \). Additionally, atomic numbers are incorporated as weighted features:
\begin{equation} \label{eq:atomic_numbers}
\mathbf{z} = [x_A Z_A, x_B Z_B, x_C Z_C, \dots]
\end{equation}
where \( Z_i \) is the atomic number of element \( i \). To capture interatomic effects, second-order (pairwise) and third-order (triplet) interaction terms are introduced:
\begin{equation} \label{eq:pairwise}
x_{ij} = x_i x_j, \quad x_{ijk} = x_i x_j x_k
\end{equation}
for all unique pairs and triplets of elements. The final feature set consists of the original concentrations, weighted atomic numbers, and interaction terms:
\begin{equation} \label{eq:features}
\mathbf{X} = \begin{array}{l}
[x_A, x_B, x_C, \dots, \\
x_A Z_A, x_B Z_B, x_C Z_C, \dots, \\
x_{AB}, x_{AC}, x_{BC}, \dots, \\
x_{ABC}, x_{ABD}, \dots]
\end{array}
\end{equation}

The error inherent in DFT calculations (that has its origin from the approximation used for the exchange and correlation functional) can be quantified as the difference between the experimental and theoretical determined enthalpy of formation. Hence, we introduce the term $H_{\text{corr}}$ as 

\begin{equation} \label{eq:error}
H_{\text{corr}}=H_{\text{DFT}} - H_{\text{expt}},
\end{equation}
and we strive here to use machine learning algorithms to make good estimates of $H_{\text{corr}}$ when experimental data ($H_{\text{expt}}$) are missing. 
For a simple linear model, the predicted enthalpy correction \( H_{\text{corr}} \) is obtained as a linear combination of the features \( \mathbf{X} \) and the model parameters \( \theta \), which include the weight coefficients \( w_i \) and the bias term \( b \), extracted via a standard least-squares fit:  
\begin{equation} \label{eq:H_corr_linear}
H_{\text{corr}} = \begin{array}{l}
b + w_1 x_A + w_2 x_B + w_3 x_C + \dots, \\
w_4 (x_A Z_A) + w_5 (x_B Z_B) + w_6 (x_C Z_C) + \dots, \\
w_7 x_{AB} + w_8 x_{AC} + w_9 x_{BC} + \dots, \\
w_{10} x_{ABC} + w_{11} x_{ABD} + \dots
\end{array}
\end{equation}
This can be expressed more compactly in matrix notation:
\begin{equation} \label{eq:H_corr_matrix}
H_{\text{corr}} = \mathbf{w}^T \mathbf{X} + b
\end{equation}
where:
- \( \mathbf{w} \) is the vector of weight coefficients,
- \( \mathbf{X} \) is the vector of input features,
- \( b \) is the bias term. Results from this simplistic approach to estimating \( H_{\text{corr}} \) are analyzed below and compared to data obtained from more advanced ML algorithms that undergo supervised training. The details of one such ML method are described below.

A neural network model has been implemented as a multi-layer perceptron (MLP) regressor with three hidden layers containing up to 250, 150, and 100 neurons, respectively. The predicted enthalpy correction, \( H_{\text{corr}} \), as defined in Eqn.~\ref{eq:error}, is obtained as:
\begin{equation} \label{eq:H_corr}
H_{\text{corr}} = f(\mathbf{X}, \theta),
\end{equation}
where \( f \) represents the neural network function with learnable network parameters \( \theta \). 
We investigate here if a neural network can result in values of $H_{\text{corr}}$ as given by Eqn.\ref{eq:H_corr} that capture the values given in Eqn.~\ref{eq:error}, and if such a neural network can make accurate predictions of $H_{\text{corr}}$  when experimental data are missing. 
The total DFT corrected enthalpy is given by
\begin{equation} \label{eq:H_pred}
H_{\text{pred}} = H_{\text{DFT}} -H_{\text{corr}},
\end{equation}
where \( H_{\text{DFT}} \) is the enthalpy from DFT calculations. As is demonstrated below, a neural network that is trained for certain concentrations of a ternary system where $H_{\text{expt}}$ is known, can give reliable values for $H_{\text{pred}}$ even for concentrations where $H_{\text{expt}}$ is missing.  

Overfitting in the training steps has been controlled through several strategies: 1. leave-one-out cross-validation (LOOCV) that ensures that each data point is tested individually, preventing memorization of training data; 2. feature selection that ensures  avoidance redundant descriptors; and 3. early stopping that prevents excessive weight updates once validation performance stabilizes, avoiding unnecessary complexity.

The model's predictive performance has been evaluated using the root-mean-square error (RMSE) across both LOOCV and k-fold cross-validation (in this work we used five folds). The final trained model, along with feature scaling parameters, has been saved for future predictions. This approach enhances the accuracy of computed formation enthalpies while maintaining interpretability in terms of elemental interactions, providing a physics-informed correction to DFT calculations for alloy thermodynamics.

\section{Results and discussion}

We have performed DFT calculations, as described above, for all the known enthalpies of formation of the two ternary phase diagrams, for Al-Ni-Pd and Al-Ni-Ti, at ambient conditions, resulting in a total of 34 systems. These systems were randomly divided into a ML training set, with approximately 3/4 of the data used for model training, and a test set, consisting of about 1/4 of the data. The latter data were used for assessing the ability of the trained neural network to make accurate predictions of \( H_{\text{corr}} \), as will be discussed below.  

\begin{figure}[ht!]
\begin{center}
\includegraphics[width=0.47\textwidth,{angle=180}, trim={39mm 0 39mm 0},clip]{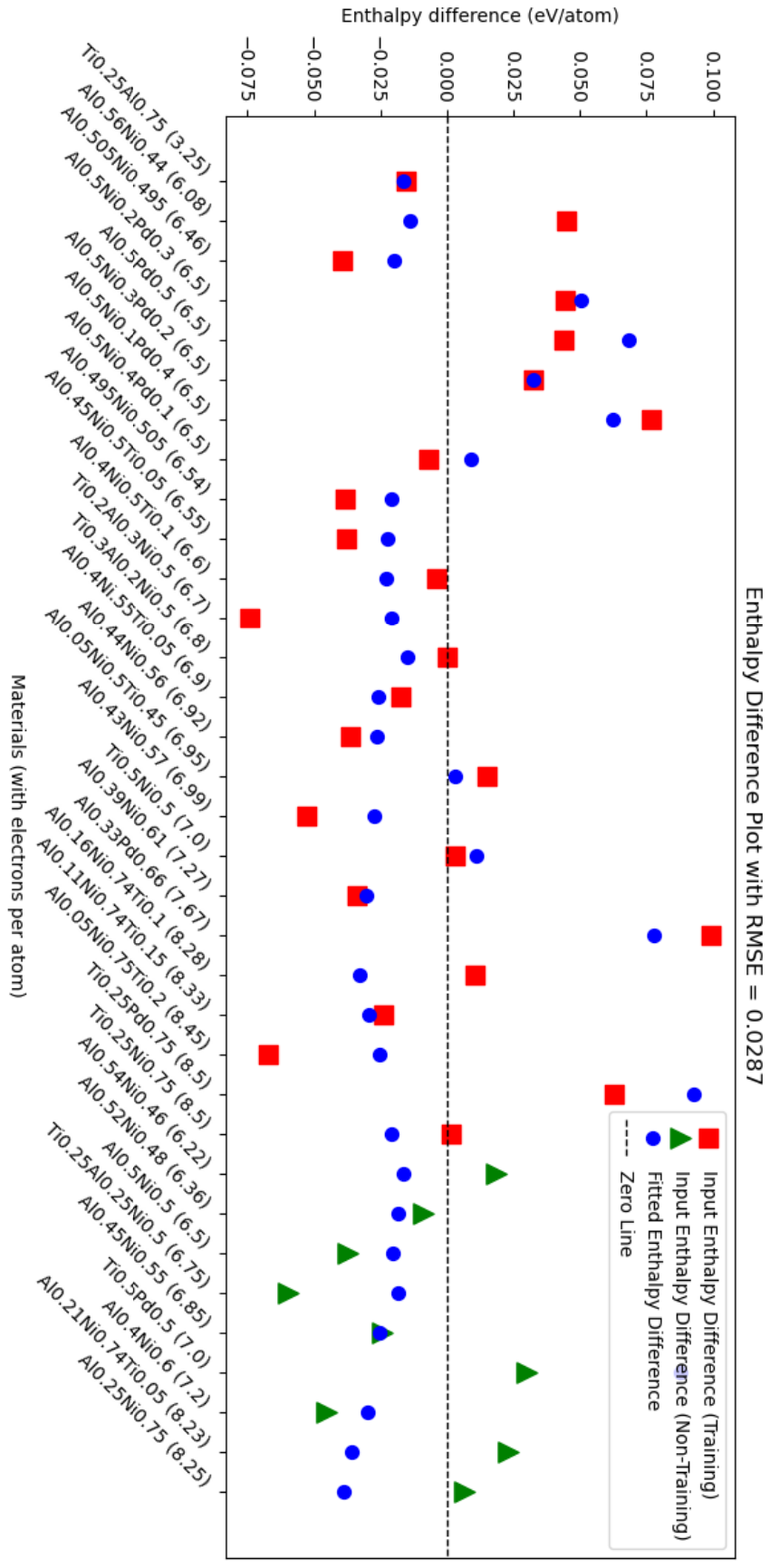}
\caption{Values of $H_{\text{corr}}$ obtained from Eqn.~\ref{eq:error} (red squares and green triangles) and from Eq.\ref{eq:H_corr_linear}, the linear model discussed in the text (blue dots), for all systems investigated in this study. The compounds are listed according to number of valence electrons (number given in parenthesis to the right of each chemical formula). The red squares have been used in the training set and the green triangles are used in the test set.  
The resulting RMSE for all the systems in the figure (both from the training set and not) is 28.7 meV/atom. Separated RMSEs for the training and prediction sets are 24.9 meV/atom and 31.4 meV/atom, respectively. }
\label{linear_model}
\end{center}
\end{figure}

Fig.~\ref{linear_model} illustrates the performance of the linear regression model, and how it reproduces the values of $H_{\text{corr}}$, as defined in Eqn.~\ref{eq:error}.
The compounds and alloys are grouped by their respective training and prediction sets, with each set further grouped according to their average valence electron count.

While the linear model captures some of the systematic errors present in the raw data, its overall performance remains limited. A key observation from Fig.~\ref{linear_model} is the significant dispersion of $H_{\text{corr}}$ from Eqn.~\ref{eq:error}, with no clear trend linking them to elemental composition or electron count. This suggests that even with the inclusion of cross-terms and higher-order interactions (such as \( x_A x_B \) and \( x_A x_B x_C \)), the linear model struggles to fully describe the complex energy corrections required to align the DFT results with experimental values.  

Despite its relatively simple structure, the linear model does provide partial improvements in some cases, but its effectiveness varies considerably across different materials. Certain structures exhibit moderate reductions in error, while others remain largely unaffected. The lack of a systematic pattern in these results highlights the non-trivial nature of the underlying enthalpy corrections—suggesting that while some deviations may be approximated by a combination of concentration and atomic number terms, many others arise from interactions that are not easily captured in a linear framework. This is particularly evident for materials where the error remains large despite the inclusion of all terms in Eqn.\ref{eq:H_corr_linear}, indicating that important nonlinear effects are still missing. 
The error of the training set is on the order of 25 meV/atom, while that of the test set is on the order of 31 meV/atom. These numbers should be compared to typical experimental error bars, as even for high-quality calorimetric measurements of alloys, the precision can be on the order of kJ/mol, which corresponds to approximately 17 meV/atom \cite{exp_accuracy}.

Next, we discuss the results of the neural network model and how training of such a network significantly improves the accuracy of $H_{\text{corr}}$ from Eqn.~\ref{eq:H_corr}.
To ensure the reliability of our network model, we systematically examined the convergence of the model by incrementally increasing the number of systems in the training set, starting from as few as five structures. The RMSE of LOOCV initially rises, reaching a peak of approximately 40 meV/atom (data not shown). This suggests that certain key structures must be included in the training set to achieve better accuracy. Beyond this point, as additional systems are incorporated, the RMSE steadily decreases, indicating convergence.  

\begin{figure}[!ht]
  \centering   \includegraphics[width=0.5\textwidth, trim={0 0 0 8cm},clip]{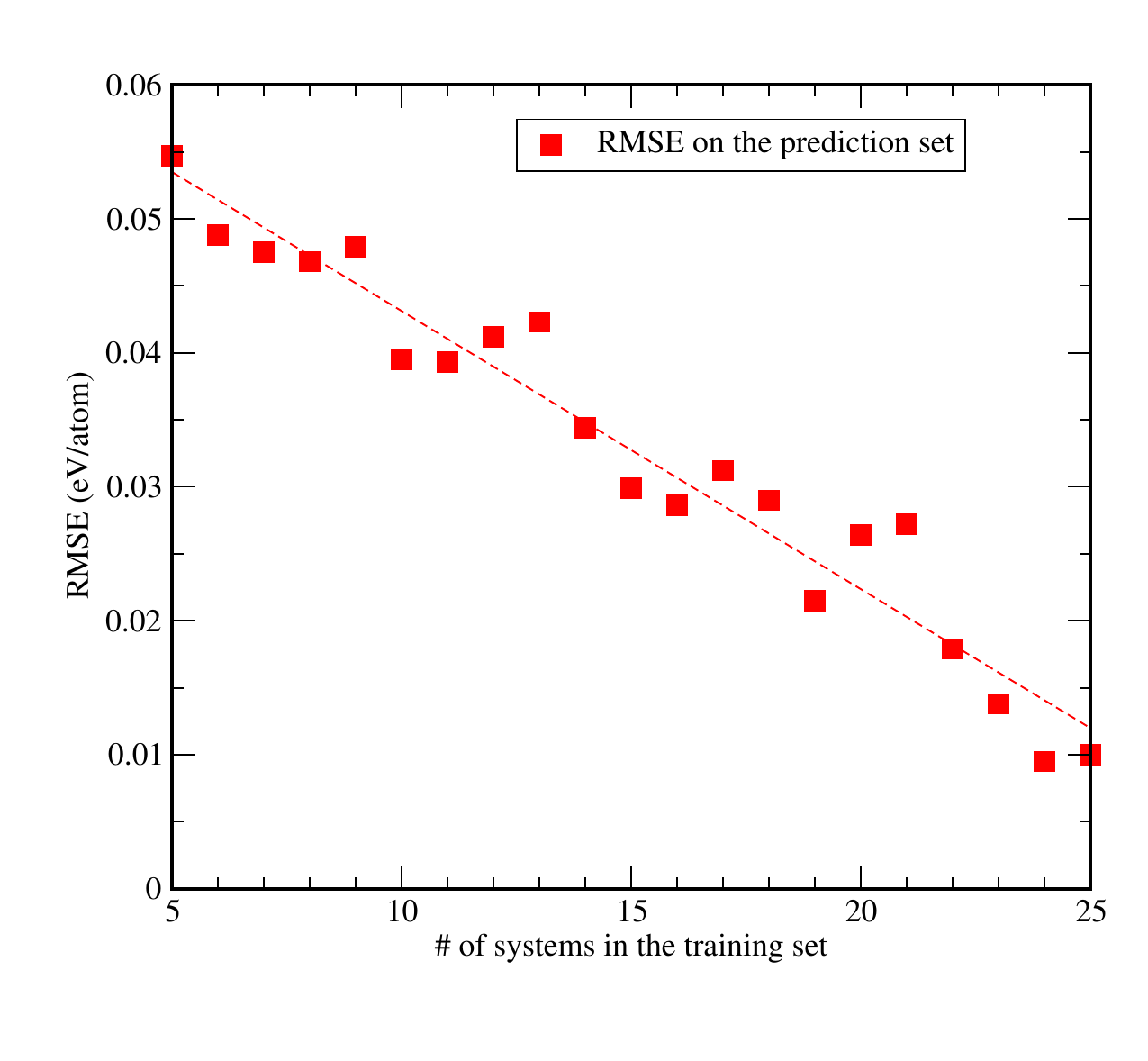}
    \caption{Convergence of RMSE on the prediction set with the number of systems in the training set}
    \label{RMSE}
\end{figure}

This trend is even more evident when evaluating the RMSE on the test (prediction) set, which is never part of the training process. As shown in Fig.~\ref{RMSE}, 
a clear decrease in RMSE of this set of systems is observed as more materials are added to the training set. With approximately 25 structures in the training set, the RMSE on the prediction set falls down to 10 meV/atom, demonstrating a substantial improvement in predictive accuracy. 

\begin{figure}[ht!]
\begin{center}
\includegraphics[width=0.47\textwidth,{angle=180}, trim={39mm 0 39mm 0},clip]{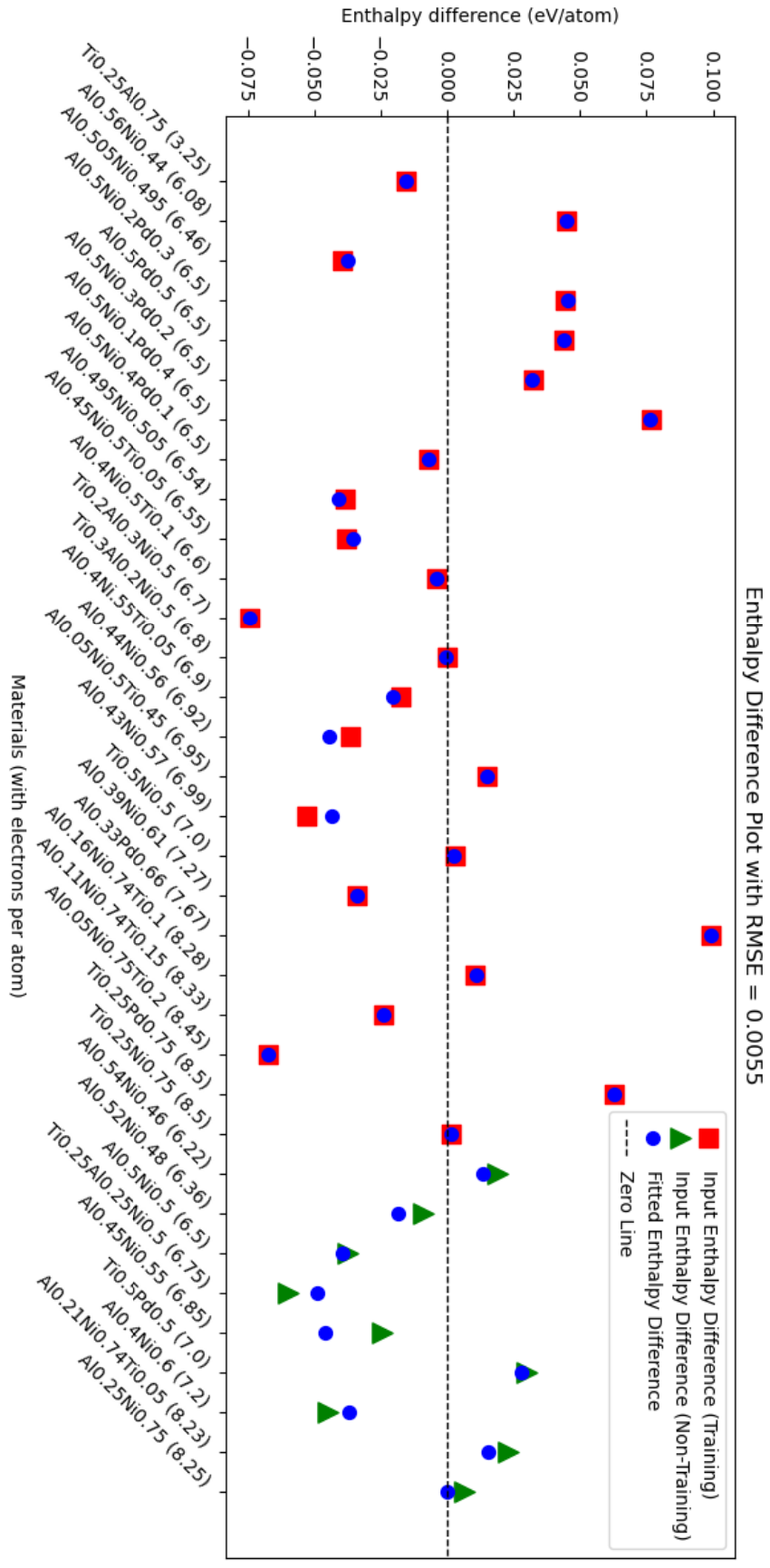}
\caption{Values of $H_{\text{corr}}$ obtained from Eqn.~\ref{eq:error} (red squares and green triangles) and from Eqn.~\ref{eq:H_corr_linear}, the trained neural network (blue dots), for all systems investigated in this study. The compounds are listed according to number of valence electrons (number given in parenthesis to the right of each chemical formula). The red squares have been used in the training set and the green triangles are used in the test set.   The resulting RMSE for all the systems in the figure (both from the training set and not) is 5.5 meV/atom. Separated RMSEs for the training and prediction sets are 2.7 meV/atom and 10.6 meV/atom, respectively. }
\label{neural_model}
\end{center}
\end{figure}

The performance of the most accurately trained neural network to reproduce the experimental values of $H_{\text{corr}}$  is shown in Fig.~\ref{neural_model}. Here we compare $H_{\text{corr}}$ from Eqns.~\ref{eq:error} and \ref{eq:H_corr}, both for data-points in the training set and outside of it. 
As seen, this model provides a more flexible and expressive correction, significantly reducing the difference between $H_{\text{corr}}$ obtained from Eqn.~\ref{eq:error} and Eqn.~\ref{eq:H_corr}.
This improvement in capturing the true values of $H_{\text{corr}}$ (Eqn.~\ref{eq:error}) reflects the model’s superior ability to capture the complex relationship between elemental composition and enthalpy deviations. The error in DFT formation enthalpies is inherently structured but highly non-trivial, requiring a model capable of learning intricate dependencies beyond simple interaction terms. The neural network excels in this aspect, successfully identifying patterns that the linear model (Fig.~\ref{linear_model}) struggles to represent. 

The ability of the neural network to predict values of $H_{\text{corr}}$ for data points outside the training set is particularly demonstrated in Fig.~\ref{neural_model}. Note that we compare values of $H_{\text{corr}}$ obtained from Eqn.~\ref{eq:H_corr} for systems outside the training set (blue dots) to exact values obtained from Eqn.~\ref{eq:error} (green triangles). The good agreement between the two sets of data points shows that the neural network considered here can accurately estimate how DFT-based calculations should be corrected to obtain an accurate enthalpy of formation, even for systems where no experimental data is available. 
To be more quantitative, we note that the RMSE of the difference between $H_{\text{corr}}$ from experimental and neural network-generated data (Eqn.~6 and Eqn.~10, respectively) is 2.7 meV/atom for the training set. The corresponding value for the test set is 10.6 meV/atom. This demonstrates that for the considered test set of systems, DFT-calculated heat of formation values (which we estimate to have an RMSE of approximately 41 meV/atom) can be significantly improved using the approach presented here, reducing the RMSE to approximately 10 meV/atom without relying on any experimental input.

\begin{figure}[!ht]
  \centering   \includegraphics[width=0.5\textwidth, trim={0 0 0 11cm},clip]{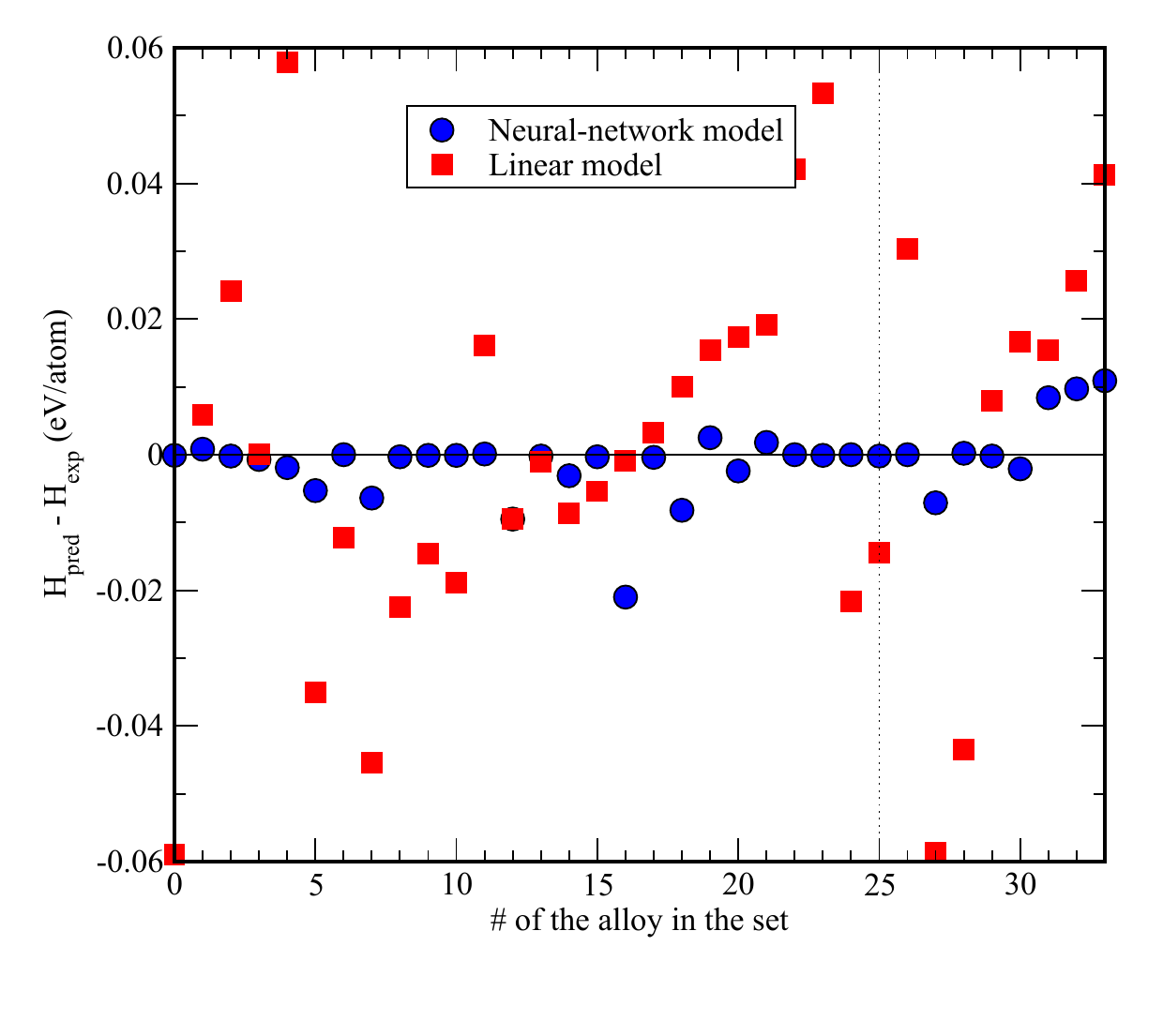}
    \caption{Difference between experimental heat of formation and the values obtained by correcting DFT data with the linear model (Eqn.~\ref{eq:H_pred}, see text) using the linear model (Eqn.~\ref{eq:H_corr_linear}, data points as red squares) and the neural network (Eqn.~\ref{eq:H_corr}, data points as blue dots).}
    \label{diff_model}
\end{figure}

A direct comparison of the two approaches analyzed in this paper is shown in Fig.~\ref{diff_model}, 
which highlights the difference between experimental data and the DFT corrected values of $H$, as obtained from the linear and neural network models. Nearly all points of the ML model exhibit a reduction in error compared to the linear model, confirming that the machine learning approach better captures the errors of the DFT calculations. The largest improvements occur in compositions where the linear model performed particularly poorly, reinforcing the idea that these deviations arise from multi-body interactions and electronic effects that cannot be approximated through additive corrections alone. However, it is also noteworthy that certain structures still show non-negligible residual errors, even with the neural network model. This suggests that additional factors—such as temperature-dependent phase behavior, local electronic configurations, or unaccounted-for experimental uncertainties may contribute to the remaining discrepancies. For completeness, we list in Table~\ref{tab:data} all experimental values of the heat of formation used in this investigation, together with the predicted values using Eqns.9 and 10. The Table also shows the difference between predicted and measured values of the heat of formation together with the difference between DFT calculated results and experimental data. 

In summary, while the linear model offers some level of correction, its predictive power is ultimately constrained by its functional form. Even with interaction terms such as \( x_A x_B \) and \( x_A x_B x_C \), it cannot fully account for the intricate relationships governing enthalpy corrections. The neural network, in contrast, demonstrates a much greater ability to capture these relationships, leading to a substantial improvement in predictive accuracy.  

We also observe that larger errors 
often signal potential issues with experimental measurements or limitations in the initial DFT model. This is particularly evident in the Al-Ni system, where alloys with nearly identical compositions exhibit unexpectedly large variations in their DFT-experiment discrepancies. For example, Al\textsubscript{0.5}Ni\textsubscript{0.5} shows a small deviation of -2.1 meV/atom, while Al\textsubscript{0.52}Ni\textsubscript{0.48} exhibits a much larger deviation of -9.5 meV/atom, despite only a 2 at.\% difference in composition. Similarly, Al\textsubscript{0.54}Ni\textsubscript{0.46} has a deviation of -5.3 meV/atom, while Al\textsubscript{0.56}Ni\textsubscript{0.44} returns to a much lower deviation of -0.1 meV/atom. The reason for this fluctuation remains unclear, as one would expect a smooth variation in enthalpy differences with concentration. These discrepancies may arise from subtle electronic structure effects, experimental uncertainties, or limitations in the theoretical model that are not fully captured even with machine learning corrections. Whatever the reason, the persistence of such deviations highlights the complexity of phase stability predictions and the need for further refinement of both theoretical and experimental approaches. This underscores the necessity of advanced modeling techniques for improving DFT-based formation enthalpies, particularly in multicomponent systems where small energy differences dictate phase stability.  

\section{Conclusion}
In this work, we have investigated a machine learning-based approach and its ability to enhance the accuracy of density functional theory (DFT) calculations for alloy formation enthalpies, particularly in ternary phase stability calculations. By utilizing a neural network model trained to predict the discrepancies between DFT-calculated and experimentally measured enthalpies of formation, we have significantly reduced the intrinsic energy error of DFT based calculations, an error that is the key limiting factor for reliable predictions of the phase stability of complex systems, such as binary- and ternary compounds and alloys.

The neural network model, which incorporates a structured feature set with elemental concentrations, atomic numbers, and interaction terms, is shown here to be able to significantly reduce the error compared to a simple linear correction model. While the linear model could capture about 25\% of the error, the machine learning approach drastically improved both the data points resulting from the training step of the study, as well as the the predictive power when applied to a test set of systems, that were not included in the training step. The approach suggested here hence leads to much more reliable predictions of the enthalpy of formation, which is key when comparing the energy of competing phases and the determination of binary and ternary phase diagrams.  

When applied to the Al-Ni-Pd and Al-Ni-Ti ternary alloy systems, we observed that the machine learning model not only reduced the root-mean-square error (RMSE) of theory, but also revealed that larger errors in some cases were indicative of potential issues with experimental measurements or the initial model. This was particularly evident in alloys with similar concentrations, where discrepancies between DFT and experiment were most pronounced.

Overall, the work presented here demonstrate that by integrating machine learning methods with first-principles calculations, forming a method we refer to as as Error Corrected Density Functional Theory (EC-DFT), one can significantly improve the accuracy of phase stability predictions, making them more reliable for practical applications. 
The methodology presented here provides a scalable and transferable framework for enhancing the predictive power of DFT while maintaining interpretability in terms of elemental interactions. This approach has the potential to accelerate computational materials design and aid in the development of advanced materials with optimized properties for various applications.

\section*{Acknowledgements}
S.I.S. acknowledge the support from the Swedish Research Council (VR) (Grant No. 2023-05247).
E. K. D.-Cz. and O.E. acknowledge support from the Wallenberg Initiative Materials Science for Sustainability (WISE) funded by the Knut and Alice Wallenberg Foundation (KAW), STandUPP, eSSENCE, and NL-ECO: Netherlands Initiative for Energy-Efficient Computing (with project number NWA. 1389.20.140) of the NWA research program.
O.E. acknowledges financial support from the Swedish Research Council (VR), the European Research Council through the ERC Synergy Grant 854843-FASTCORR and the Knut and Alice Wallenberg Foundation (KAW-Scholar program).
The computations were enabled by resources provided by the National Academic Infrastructure for Supercomputing in Sweden (NAISS), partially funded by the Swedish Research Council through grant agreement no. 2022-06725.

\section*{Declaration of Competing Interest}
The authors declare that they have no known competing financial interests or personal relationships that could have appeared to influence the work reported in this article.


\section*{Appendix - Table~\ref{tab:data}}

\begin{table*}[]
\caption{Number of valence electrons ($n_{\text{v.e.}}$) for the investigated alloys and compounds (Composition) at their experimentally reported structures indicated by their Space Group. Measured heat of formation ($H_{\text{expt}}$), predicted heat of formation ($H_{\text{pred}}$), the deviation between the predicted and measured enthalpies of formation ($H_{\text{pred}}-H_{\text{expt}}$) and the deviation between the DFT estimated and measured data ($H_{\text{DFT}}-H_{\text{expt}}$). All energies are in units of eV/atom. If not stated otherwise, the experimental heat of formation is from the database of Ref.~\cite{exp_accuracy}.}
\centering
\begin{tabular} {ccllllrr} 
\toprule
nr& $n_{v.e.}$&Composition & Space Group & $H_{\text{expt}}$ & $H_{\text{pred}}$& $H_{\text{pred}}-H_{\text{expt}}$&$H_{\text{DFT}}-H_{\text{expt}}$\\
\midrule
1       &3.25   &Ti$_{0.25}$Al$_{0.75}$                 &I4/mmm &-0.3793        &-0.3791& 0.0002        &-0.0153 \\
2       &6.08   &Al$_{0.56}$Ni$_{0.44}$                 &Pm-3m  &-0.6022$^a$    &-0.6021& 0.0001        & 0.0450 \\
3       &6.47   &Al$_{0.505}$Ni$_{0.495}$               &Pm-3m  &-0.6291$^a$    &-0.6309&-0.0018        &-0.0392 \\
4       &6.50   &Al$_{0.50}$Ni$_{0.20}$Pd$_{0.30}$      &Pm-3m  &-0.8080        &-0.8088&-0.0008        & 0.0444 \\
5       &6.50   &Al$_{0.50}$Pd$_{0.50}$                 &Pm-3m  &-0.9457        &-0.9455& 0.0002        & 0.0440 \\
6       &6.50   &Al$_{0.50}$Ni$_{0.30}$Pd$_{0.20}$      &Pm-3m  &-0.7476        &-0.7469& 0.0007        & 0.0324 \\
7       &6.50   &Al$_{0.50}$Ni$_{0.10}$Pd$_{0.40}$      &Pm-3m  &-0.9033        &-0.9031& 0.0002        & 0.0766 \\
8       &6.50   &Al$_{0.50}$Ni$_{0.40}$Pd$_{0.10}$      &Pm-3m  &-0.6777        &-0.6778&-0.0001        &-0.0072 \\
9       &6.54   &Al$_{0.495}$Ni$_{0.505}$               &Pm-3m  &-0.6343$^a$    &-0.6319& 0.0024        &-0.0385 \\
10      &6.55   &Al$_{0.45}$Ni$_{0.50}$Ti$_{0.05}$      &Pm-3m  &-0.6156        &-0.6181&-0.0025        &-0.0378 \\
11      &6.60   &Al$_{0.40}$Ni$_{0.50}$Ti$_{0.10}$      &Pm-3m  &-0.6229        &-0.6228& 0.0001        &-0.0041 \\
12      &6.70   &Ti$_{0.20}$Al$_{0.30}$Ni$_{0.50}$      &Fm-3m  &-0.5599        &-0.5598& 0.0001        &-0.0744 \\
13      &6.80   &Ti$_{0.30}$Al$_{0.20}$Ni$_{0.50}$      &Fm-3m  &-0.5690        &-0.5689& 0.0001        &-0.0002 \\
14      &6.90   &Al$_{0.40}$Ni$_{0.55}$Ti$_{0.05}$      &Pm-3m  &-0.5882        &-0.5851& 0.0031        &-0.0175 \\
15      &6.92   &Al$_{0.44}$Ni$_{0.56}$                 &Pm-3m  &-0.5835$^a$    &-0.5753& 0.0082        &-0.0364 \\
16      &6.95   &Al$_{0.05}$Ni$_{0.50}$Ti$_{0.45}$      &Pm-3m  &-0.4218        &-0.4218& 0.0000        & 0.0150 \\
17      &6.99   &Al$_{0.43}$Ni$_{0.57}$                 &Pm-3m  &-0.5566$^a$    &-0.5663&-0.0097        &-0.0530 \\
18      &7.00   &Ti$_{0.50}$Ni$_{0.50}$                 &Pm-3m  &-0.3749        &-0.3747& 0.0002        & 0.0028 \\
19      &7.27   &Al$_{0.39}$Ni$_{0.61}$                 &Pm-3m  &-0.5317$^a$    &-0.5313& 0.0004        &-0.0337 \\
20      &7.67   &Al$_{0.33}$Pd$_{0.66}$                 &Pnma   &-0.9048        &-0.9048& 0.0000        & 0.0993 \\
21      &8.28   &Al$_{0.16}$Ni$_{0.74}$Ti$_{0.10}$      &Pm-3m  &-0.4404$^b$    &-0.4406&-0.0002        & 0.0107 \\
22      &8.33   &Al$_{0.11}$Ni$_{0.74}$Ti$_{0.15}$      &Pm-3m  &-0.4102        &-0.4099& 0.0003        &-0.0238 \\
23      &8.45   &Al$_{0.05}$Ni$_{0.75}$Ti$_{0.20}$      &P63/mmc&-0.3616        &-0.3616& 0.0000        &-0.0674 \\
24      &8.50   &Ti$_{0.25}$Pd$_{0.75}$                 &P63/mmc&-0.6737        &-0.6737& 0.0000        & 0.0626 \\
25      &8.50   &Ti$_{0.25}$Ni$_{0.75}$                 &P63/mmc&-0.4374$^c$    &-0.4371& 0.0003        & 0.0016 \\
26      &6.22   &Al$_{0.54}$Ni$_{0.46}$                 &Pm-3m  &-0.6177        &-0.6124& 0.0053        & 0.0187 \\
27      &6.36   &Al$_{0.52}$Ni$_{0.48}$                 &Pm-3m  &-0.6301$^a$    &-0.6206& 0.0095        &-0.0090 \\
28      &6.50   &Al$_{0.50}$Ni$_{0.50}$                 &Pm-3m  &-0.6405$^a$    &-0.6384& 0.0021        &-0.0372 \\
29      &6.75   &Ti$_{0.25}$Al$_{0.25}$Ni$_{0.50}$      &Fm-3m  &-0.5783        &-0.5892&-0.0109        &-0.0598 \\
30      &6.85   &Al$_{0.45}$Ni$_{0.55}$                 &Pm-3m  &-0.6053        &-0.5843& 0.0210        &-0.0246 \\
31      &7.00   &Ti$_{0.50}$Pd$_{0.50}$                 &Pmma   &-0.5524        &-0.5505& 0.0019        & 0.0299 \\
32      &7.20   &Al$_{0.40}$Ni$_{0.60}$                 &Pm-3m  &-0.5317        &-0.5401&-0.0084        &-0.0451 \\
33      &8.23   &Al$_{0.21}$Ni$_{0.74}$Ti$_{0.05}$      &Pm-3m  &-0.4456$^c$    &-0.4385& 0.0071        & 0.0228 \\
34      &8.25   &Al$_{0.25}$Ni$_{0.75}$                 &Pm-3m  &-0.4198        &-0.4134& 0.0064        & 0.0065 \\
\bottomrule
{$^{a}$Ref.~\cite{NASH2001228}}& $^b$Ref.~\cite{AlNiTi_hf}& $^c$Ref.~\cite{GUO1998181}&&&&\\
\end{tabular}
\label{tab:data}
\end{table*}

\printcredits
\bibliographystyle{elsarticle-num} 
\bibliography{ref.bib}

\end{document}